# Counter-propagating frequency mixing with Terahertz waves in diamond

Matteo Clerici,[1,2,*] Lucia Caspani,[1] Eleonora Rubino,[1,3] Marco Peccianti,[4] Marco Cassataro,[1,5] Alessandro Busacca,[5] Tsuneyuki Ozaki,[1] Daniele Faccio,[2] and Roberto Morandotti[1]

[1]*INRS-EMT, 1650 Blvd. Lionel-Boulet, Varennes, Québec J3X 1S2, Canada.*
[2]*School of Engineering and Physical Sciences, Heriot-Watt University, SUPA, Edinburgh EH14 4AS, UK.*
[3]*Dipartimento di Scienza e Alta Tecnologia, Università degli Studi dell'Insubria, via Valleggio 11, 22100 Como, Italy.*
[4]*Institute for complex Systems-CNR via dei Taurini 19, 00185 Roma, Italy.*
[5]*DIEET, Università di Palermo, viale delle Scienze 9, 61-90133 Palermo, Italy.*
*\*Corresponding author: clerici@emt.inrs.ca*

Frequency conversion by means of Kerr-nonlinearity is one of the most common and exploited nonlinear optical processes in the UV, visible, IR and Mid-IR spectral regions. Here we show that wave mixing of an optical field and a Terahertz wave can be achieved in diamond, resulting in the frequency conversion of the THz radiation either by sum- or difference-frequency generation. In the latter case, we show that this process is phase-matched and most efficient in a counter-propagating geometry.

Terahertz (THz) radiation covers the spectral range between 0.1 and 10 THz (3 mm – 30 µm) and is gathering an increasing interest both for spectroscopic applications and as a playground for fundamental studies *e.g.* on nonlinear and extreme-nonlinear optical effects [1–5]. Furthermore, THz radiation is also attracting attention for its possible application *e.g.* as a control field for integrated nonlinear optics [6].

Although several studies have already investigated the wave mixing of THz and optical fields via the Kerr $-\chi^{(3)}-$ nonlinearity, especially in gases (see *e.g.* [7,8]), only a few have addressed the possibility of performing nonlinear wave mixing in bulk samples, typically exploiting electric-field-induced second harmonic generation [9–11] and more recently, four-wave mixing for THz wave generation [12,13].

In this Letter, we report on the wave mixing of THz and near-infrared radiation in a <100>-cut diamond bulk sample. We show that two processes, namely sum-frequency (*SF*) and difference-frequency generation (*DF*) coexist, and that counter-propagating *DF*, *i.e.* taking place for an optical pulse interacting with a counter-propagating THz field, appears to be the most efficient process thanks to the longer coherence length.

We start by considering the *SF* and *DF* interactions:

$$SF: \quad 2\omega_p + \omega_T = \omega_{SF}$$
$$DF: \quad 2\omega_p = \omega_T + \omega_{DF} \quad , \quad (1)$$

where $\omega_p$ is the optical pump frequency (in our case corresponding to a 792 nm wavelength), $\omega_T$ is the seed THz field carrier frequency, and $\omega_{SF}/\omega_{DF}$ is the frequency of the idler wave resulting from the *SF/DF* process (from hereon we shall refer to *SF/DF* for both the effect and the generated field). Considering the case of a collinear interaction of plane, monochromatic waves, the phase-matching condition reduces to a scalar equation for the involved wavevectors. In this case, for both processes two different configurations are possible:

$$\begin{aligned}
SF\text{-}P: &\quad k_{SF} = 2k_p + k_T &\rightarrow&\quad \Delta k_{SF-P} = k_{SF} - 2k_p - k_T \\
SF\text{-}C: &\quad k_{SF} = 2k_p - k_T &\rightarrow&\quad \Delta k_{SF-C} = k_{SF} - 2k_p + k_T \\
DF\text{-}P: &\quad k_{DF} + k_T = 2k_p &\rightarrow&\quad \Delta k_{DF-P} = k_{DF} - 2k_p + k_T \\
DF\text{-}C: &\quad k_{DF} - k_T = 2k_p &\rightarrow&\quad \Delta k_{DF-C} = k_{DF} - 2k_p - k_T
\end{aligned} \quad , (2)$$

where $k_i$ denotes the *i-th* field wavevector (*i* being *p* for the pump, *T* for the THz field, with *SF* and *DF* as stated above) whereas *P* and *C* indicate the co- and counter-propagating configurations, respectively, and $\Delta k$ is the phase mismatch.

It is worth noting that in a dispersive medium the phase mismatch $\Delta k$ for the four interactions in Eq. (2) is different since $k_{SF} \neq k_{DF}$, as the two frequencies $\omega_{SF}$ and $\omega_{DF}$ are different.

In our experimental configuration, the THz pulse is generated by laser-induced plasma and shows peak electric fields in the order of few MV/cm with a duration of 90 fs (full-width at half maximum). The details of the source are reported elsewhere [14]. The instantaneous electric field and bandwidth, measured by Air-Biased Coherent Detection [15] are shown in Fig. 1. The optical pump, delivered by a Ti:Sapphire amplifier, has a duration of 60 fs (full width at half maximum) and carrier wavelength of 792 nm. The frequencies of the *SF* and *DF* fields [$\omega_{SF}$ and $\omega_{DF}$ in Eq. (1)] corresponding to the 0-25 THz seed bandwidth are overlaid in Fig. 1(b) – right scale.

In order to observe the nonlinear wave mixing between a THz and an optical pulse, a suitable material featuring low absorption at all the wavelengths involved in the process is essential. Diamond is the perfect candidate showing negligible absorption (<1 cm$^{-1}$) both in the THz and in the far-infrared bandwidth, as well as at 800 nm and at the $\lambda_{SF/DF}$ wavelength (~400 nm) [16]. Furthermore, the high nonlinear coefficient guarantees reasonable frequency conversion efficiencies[17–19].

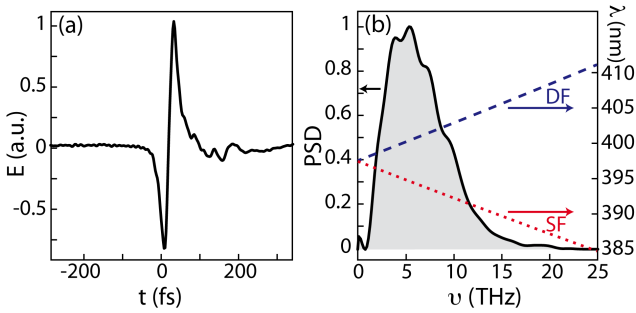

Fig. 1. (Color online) Instantaneous electric field (a) and power spectral density (b) for the THz pulse employed in our experiments. In (b), the dashed (blue) and dotted (red) lines show the *SF* and *DF* wavelengths corresponding to the THz seed frequency, respectively (right scale).

In our experiment, we have employed four different diamond samples. Two were single crystal CVD slabs (*Element Six TM*) of 500 µm and 300 µm thickness (4.5 x 4.5 mm and 3 x 3 mm aperture, respectively). The others were two thinner polycrystalline films (*Diamond Materials GmbH*) of 100 µm and 50 µm thickness (5 mm clear aperture). In the first measurement we investigated the co-propagating geometry by overlapping, in the different diamond samples, the focused THz beam (~90 µm Gaussian beam waist) and a collimated 792 nm pump beam (1.5 mm beam waist). We hence recorded the *SF/DF* spectrum at different pump-to-THz time delays. A sketch of this configuration is shown in Fig. 2(a). Figure 2(b) shows the delay-resolved spectrum of the *DF/SF* for the 500 µm thick sample. For the collinear, co-propagating wave mixing between a 60 fs (792 nm) and a 90 fs (THz) pulse we would expect a delay-dependent signal lasting around 110 fs. On the contrary, the experimental results clearly show a trace far more extended, with a duration of nearly 8 ps [Fig. 2(b)], a value that is not consistent with the assumption of a purely co-propagating geometry.

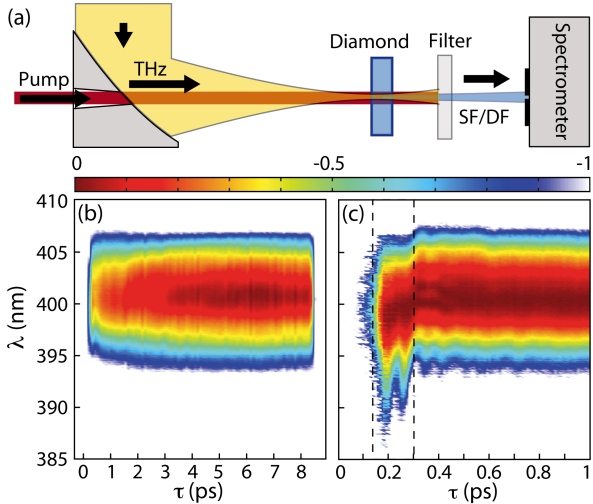

Fig. 2. (Color online) (a) Sketch of the experiment investigating the co-propagating wave mixing geometry. (b) Logarithmic representation of the wave mixing spectrogram (normalized) in the violet spectral region. (c) Zoom of (b) on the delay region where co-propagating wave mixing takes place. Note that (c) is normalized in a different way than (b) – see text for details.

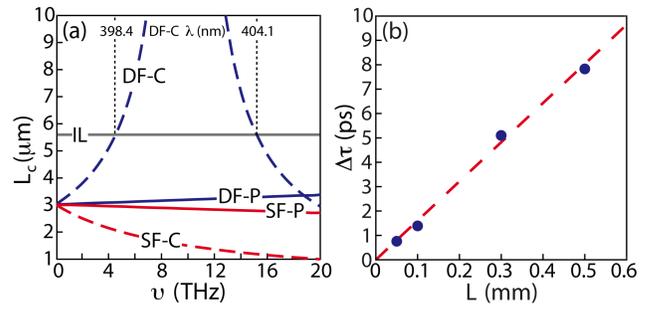

Fig. 3. (Color online) (a) Coherence lengths in function of the frequency for the co- (solid) and counter- (dashed) propagating *SF* (red) and *DF* (blue) processes. The gray horizontal line is the estimated *DF-C* interaction length (IL). The wavelength on top shows the *DF-C* bandwidth. (b) The dots show the experimentally recorded duration –$\Delta\tau$– of the *DF* signal for different crystal thicknesses L. The dashed line is the duration calculated from the pulses group velocities (see text for details).

Considering only the initial region (close to the zero-delay), we clearly observe that the recorded signal is composed of two contributions. In Fig. 2(c) we highlight this by showing the zoomed spectrum up to 1 ps delay, normalized to unit at each delay (for signals above 0.1 of the maximum recorded one, *i.e.* where the signal to noise ratio is acceptable).

The first contribution, delimited by the vertical dashed lines in Fig. 2(c), originates from the co-propagating process. For longer delays a red-shifted signal is observed lasting for much longer times. In order to understand the origin of this signal in this case, we show the coherence length $L_c \equiv \pi/|\Delta k|$ for the four different possible interaction geometries considered in Eq. (2) [Fig. 3(a)]. The solid blue and red curves are for the co-propagating *DF* and *SF*, respectively. We note that the coherence lengths of these two processes are extremely small and comparable. The *SF* and *DF* frequencies can be extracted from the dotted and dashed curves in Fig. 1(b). Their temporal phases are determined by $2\phi_P + \phi_T$ and $2\phi_P - \phi_T$, respectively, and the beating of these two signals has a component at twice the THz carrier frequency, which appears indeed at the shorter wavelengths in Fig. 2(c), as a function of the delay. The recorded *SF* component is however weaker with respect to *DF* (see also [20]).

The long-lasting, red shifted signal can thus be interpreted as the result of a more efficient backward phase matched interaction – *DF-C*, seeded by the THz signal (16.6%) reflected from the output face of the diamond sample. From a simple analysis of the coherence lengths for the counter-propagating geometries, we note that the *DF-C* is perfectly phase-matched for a 10 THz seed [dashed blue curve in Fig. 3(a)] while the *SF-C* is phase-mismatched (dashed red curve). The red shift is simply a consequence of the frequency matching shown by the blue dashed curve in Fig. 1(b) for the phase-matched THz bandwidth (around 9.9 THz, which corresponds to 401.2 nm). On the other hand, the asymmetry in the spectrum stems from the competing trends of the spectral power density peaked at ≃ 5 THz and of the phase-matching fulfilled at ≃ 10 THz.

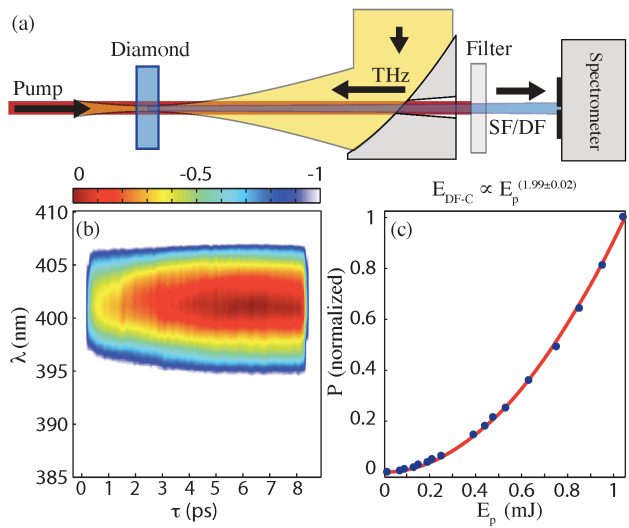

Fig. 4. (Color online) (a) Sketch of the experiment investigating the counter-propagating geometry. (b) Spectrogram of the wave mixing in the violet spectral region (log scale, normalized). (c) Normalized DF-C power as a function of the pump energy (blue dots). The red curve is a power fit with exponent 1.99 ± 0.02.

The counter-propagating phase-matching hypothesis is confirmed by the analysis of the delay-dependent DF signals recorded for the four different sample thicknesses. In a counter-propagating geometry the delay-dependent signal is expected to extend along the delay coordinate $\tau$, for $\Delta\tau \approx L(1/v_{g,T} + 1/v_{g,p})$, where $v_{g,T/p}$ are the THz and optical pulse group velocities, respectively. In our experiments, the recorded $\Delta\tau$ values for the four samples [blue dots in Fig. 3(b)] match indeed what is predicted analytically (red-dashed line). Noteworthy, no difference in the signal duration is expected between single-crystal and polycrystalline samples of equal thicknesses. The latter was verified by comparing measurements performed on either polycrystalline or single-crystal 500 μm thick diamond samples.

In order to further confirm our conclusions and to characterize the DF-C signal in the phase-matched geometry interacting with the whole input seed rather than just a reflection, we performed a second measurement directly injecting the THz pulse counter-propagating with respect to the 792 nm pump pulse in the 500 μm diamond sample [see sketch in Fig. 4(a)]. The DF-C signal is spectrally resolved for different pump-THz delays, resulting in a spectrogram [Fig. 4(b)] similar to the one measured in the previous configuration [Fig. 3(b)], except for the absence of the initial, blue-shifted part, further confirming the counter-propagating phase-matching hypothesis.

Finally, we recorded the generated DF-signal power in the counter-propagating geometry for different pump pulse energies, reported in Fig. 4(c). A power fit confirms the expected quadratic dependence. The recorded low values for the DF-signal power are mainly a consequence of the short interaction length (IL), limited by the reduced pulse overlap (~5.5 μm) due to the counter-propagating geometry: $IL = \tau_T(1/v_{g,p} - 1/v_{g,T})$ (solid gray line in Fig. 3).

In conclusion, we have shown, for the first time to the best of our knowledge, a wavelength shifting mechanism (see e.g. [21,22]) relying on a naturally phase-matched difference frequency generation process occurring in a Kerr medium (diamond) between counter-propagating waves.

Several intriguing applications can be envisaged, such as the detection and imaging of THz fields in a counter-propagating geometry. Furthermore, our results hint toward further investigations of counter-propagating wave-mixing [23].

MC acknowledges the support of the IOF People Programme (Marie Curie Actions) of the European Union's FP7-2012, KOHERENT, GA 299522. LC and ER acknowledge the support from "*Le Fonds québécois de la recherche sur la nature et les technologies*" (FQRNT)-MELS. The authors acknowledge *AXIS Photonique Inc.*, M. Bouvier and the ALLS staff for technical support; P. Di Trapani, G. Assanto and F. Légaré for enlightening discussions; B. E. Schmidt and N. Thiré for experimental support.